\documentstyle[aps,prb,twocolumn,epsf,floats]{revtex}
\draft
\begin{document}
\wideabs{

\title{Charged Excitons in a Dilute 2D Electron Gas 
       in a High Magnetic Field}

\author{
   Arkadiusz W\'ojs}
\address{
   Department of Physics, 
   University of Tennessee, Knoxville, Tennessee 37996 \\
   Institute of Physics, 
   Wroclaw University of Technology, Wroclaw 50-370, Poland}

\author{
   John J. Quinn}
\address{
   Department of Physics, 
   University of Tennessee, Knoxville, Tennessee 37996}

\author{
   Pawel Hawrylak}
\address{
   Institute for Microstructural Sciences, 
   National Research Council of Canada, Ottawa, Canada K1A 0R6}

\maketitle

\begin{abstract}
A theory of charged excitons $X^-$ in a dilute 2D electron gas in 
a high magnetic field is presented. 
In contrast to previous calculations, three bound $X^-$ states
(one singlet and two triplets) are found in a narrow and symmetric 
GaAs quantum well.
The singlet and a ``bright'' triplet are the two optically active 
states observed in experiments.
The bright triplet has the binding energy of about 1~meV, smaller 
than the singlet and a ``dark'' triplet.
The interaction of bound $X^-$'s with a dilute 2D electron gas 
is investigated using exact diagonalization techniques. 
It is found that the short-range character of the $e$--$X^-$ 
interactions effectively isolates bound $X^-$ states from a 
dilute $e$--$h$ plasma.
This results in the insensitivity of the photoluminescence 
spectrum to the filling factor $\nu$, and an exponential decrease 
of the oscillator strength of the dark triplet $X^-$ as a function 
of $\nu^{-1}$.
\end{abstract}
}

\section{Introduction}
The magneto-optical properties of quasi-two-dimensional (2D) electron 
systems have been intensively investigated experimentally\cite{%
kheng,buhmann,shields,finkelstein,glasberg,priest,hayne,nickel,%
tischler,wojtowicz,astakhov,jiang,brown} and theoretically.\cite{%
macdonald,lerner,dzyubenko1,chen,stebe,rashba,x-dot,palacios,thilagam,%
chapman,whittaker,x-fqhe,x-cf,moradi,pawel1,brum,pawel2}
For a dilute electron gas, the photoluminescence (PL) spectrum is 
determined by a charged-exciton complex $X^-$ and its interaction 
with remaining electrons.
The $X^-$ consists of two electrons and a valence hole and is similar 
to the hydrogen ion H$^-$.
Its existence in bulk semiconductors was first predicted by Lampert,
\cite{lampert} but due to small binding energy it has not been observed 
experimentally.
Stebe and Ainane\cite{stebe} showed that the binding of the second 
electron to the exciton $X$ should be enhanced in 2D systems. 
Indeed, the $X^-$ has been observed in semiconductor quantum wells 
(QW) by Kheng et al.\cite{kheng} and in many related experiments.
\cite{buhmann,shields,finkelstein,glasberg,priest,hayne,nickel,%
tischler,wojtowicz,astakhov}

The experimental observation stimulated a number of theoretical works.
\cite{chen,stebe,rashba,x-dot,palacios,thilagam,chapman,whittaker,%
x-fqhe,x-cf,moradi}
It is now well established that the only bound $X^-$ state at zero 
magnetic field is the singlet state ($X^-_s$) with the total electron 
spin $J_e=0$.
Accordingly, the PL spectrum shows only two peaks, due to the $X$ and 
$X^-_s$ recombination, split by the $X^-_s$ binding energy $\Delta_s$.
The situation is much more complicated in a magnetic field. 
In very high fields, MacDonald and Rezayi\cite{macdonald} showed that 
optically active magneto-excitons do not bind a second electron. 
They are effectively decoupled from the excess electrons due to the 
``hidden symmetry,''\cite{lerner,dzyubenko1,chen} and the PL spectrum 
is that of a single exciton, irrespective of the number of electrons 
present.

It was therefore surprising when a bound $X^-$ complex was discovered 
via numerical experiments in the lowest Landau level (LL).\cite{x-dot}
The bound complex was a triplet ($X^-_t$) with finite total angular 
momentum and a macroscopic degeneracy.
It was later shown by Palacios et al.\cite{palacios} that an isolated 
$X^-_t$ in the lowest LL has infinite radiative time $\tau_t$.
Two independent symmetries must be broken to allow for the $X^-_t$ 
recombination: the ``hidden symmetry''\cite{macdonald,lerner,dzyubenko1,%
palacios} due to an equal strength of $e$--$e$ and $e$--$h$ interactions, 
and the 2D geometrical (translational) symmetry\cite{chen,x-fqhe,dzyubenko2} 
resulting in the conservation of two angular momentum quantum numbers.
The ``hidden symmetry'' can be broken by mixing of LL's, valence band 
mixing effects, and asymmetry of the QW. 
The translational symmetry can be broken by disorder.
Therefore, the $X^-_t$ recombination probability is determined by disorder 
and scattering by additional electrons, and is expected to disappear with 
increasing magnetic field.
Also, crossing of the $X^-_t$ and $X^-_s$ PL peaks must occur at 
some value of the magnetic field, when $X^-_t$ becomes the $X^-$ 
ground state.
This hypothetical long-lived $X^-_t$ ground state in high magnetic 
fields has recently received a lot of attention.
Because the $X^-_t$ complexes carry a net charge and form LL's, they are 
expected\cite{x-fqhe,x-cf} to form (together with remaining electrons) 
the multi-component incompressible fluid states with Laughlin--Halperin
\cite{laughlin,halperin} (LH) correlations.
Since an experimental realization of such states requires reaching 
the ``hidden symmetry'' regime (long-lived $X^-_t$ ground state), an 
estimate of required magnetic fields is needed.

While variational calculations of hydrogen-like $X^-_s$ appear satisfactory,
\cite{stebe,thilagam,chapman} an accurate description of $X^-_t$ at finite 
magnetic fields is extremely difficult.
Although Whittaker and Shields\cite{whittaker} (WS) predicted a 
transition to the $X^-_t$ ground state in a GaAs/AlGaAs QW of width 
$w=10$~nm at the magnetic field of $B\approx30$~T, the experimental 
data for $B\le10$~T that was available at the time\cite{shields,%
finkelstein,glasberg} could not verify their result.
A negative answer came recently from Hayne et al.,\cite{hayne} whose
PL measurements in magnetic fields up to $50$~T seemingly precluded 
such transition.
In their spectra, $X^-_s$ remained the ground state up to 50~T, and an 
extrapolation to higher fields ruled out the singlet-triplet crossing 
at any near values.
Moreover, in clear disagreement with Ref.~\onlinecite{palacios}, strong 
$X^-_t$ PL was detected, whose intensity {\em increased} with increasing 
the magnetic field, and which at 13.5~T exceeded that of the $X^-_s$.
Results of Hayne et al.\ not only disagreed with the model of WS, but 
also suggested that a picture\cite{palacios,x-fqhe,x-cf} of long-lived 
$X^-_t$'s forming the low energy states of an $e$--$h$ plasma, worked 
out for a strictly 2D system ($w=0$) in the lowest LL, might be totally 
inadequate to realistic GaAs systems.
This suspicion was further reinforced by the unexplained lack of the 
sensitivity of PL to the filling factor of the electron gas.
The source of disagreement might be either in the description of bound 
$X^-$ states or in the description of its interaction with excess 
electrons.

In this paper we address both issues. 
We report on detailed numerical calculations of the energy and PL 
spectra of $e$--$h$ systems at high magnetic fields.
Using Lanczos-based\cite{lanczos} methods we were able to include 
in our model the effects of Coulomb interaction, LL mixing, finite 
QW width, and realistic Zeeman and cyclotron splittings.
Our calculations predict the existence of a new, optically active
bound state $X^-_{tb}$ of the triplet charged exciton.
The identification of this new state as the triplet $X^-$ state 
observed in PL explains the puzzling qualitative disagreement 
between earlier theory and experiments.
The ``bright'' $X^-_{tb}$ state is distinguished from the ``dark'' 
state $X^-_{td}$ found in earlier calculations\cite{x-dot,palacios,%
whittaker,x-fqhe,x-cf}, which is the lowest-energy triplet $X^-$ 
state at high magnetic field but remains undetected in PL experiments 
(however, see also Ref.~\onlinecite{munteanu}).
Energies and oscillator strengths of all bound complexes: $X$, $X^-_s$, 
$X^-_{tb}$, and $X^-_{td}$, are calculated as a function of the magnetic 
field and QW width.
The transition to the $X^-_{td}$ ground state at $B\approx30$~T 
is confirmed.

The interaction of $X^-$'s with additional electrons is also studied.
Because this interaction has short range, it effectively isolates the 
bound $X^-$ states from remaining electrons and only weakly affects 
PL from dilute systems, as observed by Priest et al.\cite{priest}
In particular, collisions of $X^-_{td}$ with surrounding electron gas at 
filling factors $\nu<1/5$ do not significantly enhance its oscillator 
strength.
This explains why this state is not observed in PL.

\section{Model}
In order to preserve the 2D translational symmetry of an infinite QW
in a finite-size calculation, electrons and holes are put on a surface 
of the Haldane sphere\cite{haldane1,wu} of radius $R$.
The reason to choose the spherical geometry for calculations is 
strictly technical and of no physical consequence for our results.
Because of the finite LL degeneracy, the numerical calculations on 
a sphere can be done without cutting off the Hilbert space and thus 
without breaking the 2D translational symmetry.
This allows exact resolution of the two quantum numbers conserved due 
to this symmetry: the total (${\cal L}_{\rm tot}$) and center-of-mass 
(${\cal L}_{\rm cm}$) angular momenta.
Let us note that in earlier calculations, WS\cite{whittaker} and Chapman 
et al.\cite{chapman} used planar geometry and hence could not resolve 
the ${\cal L}_{\rm cm}$ quantum number, which is essential to correctly 
identify the bound $X^-$ states and to further accurately calculate their 
energy and PL.
The exact mapping\cite{geometry} between the ${\cal L}_{\rm tot}$ and 
${\cal L}_{\rm cm}$ quantum numbers on a plane and the 2D algebra of 
the total angular momentum on a sphere (and between the respective 
Hilbert eigensubspaces) allows conversion of the results from one 
geometry to the other.
The price paid for closing the Hilbert space without breaking 
symmetries is the surface curvature which modifies interactions.
However, if the correlations to be modeled have short range that can
be described by a small characteristic length $\delta$, the effects 
of curvature are scaled by a small parameter $\delta/R$ and can be
eliminated by extrapolating the results to $R\rightarrow\infty$.
Therefore, despite all differences, the spherical geometry is equally 
well suited to modeling bound complexes as to the fractional quantum 
Hall systems (as originally used by Haldane\cite{haldane1}).

The detailed description of the Haldane sphere model can be found e.g.\ 
in Refs.~\onlinecite{haldane1,wu,geometry,fano,parentage} and (since 
it is not essential for our results) it will not be repeated here.
The magnetic field $B$ perpendicular to the surface of the sphere is 
due to a magnetic monopole placed in the center.
The monopole strength $2S$ is defined in the units of elementary flux
$\phi_0=hc/e$, so that $4\pi R^2B=2S\phi_0$ and the magnetic length is
$\lambda=R/\sqrt{S}$.
The single-particle states are the eigenstates of angular momentum $l$ 
and its projection $m$ and are called monopole harmonics.
The energies $\varepsilon$ fall into $(2l+1)$-fold degenerate angular 
momentum shells separated by the cyclotron energy $\hbar\omega_c$.
The $n$-th ($n\ge0$) shell (LL) has $l=S+n$ and thus $2S$ is a measure 
of the system size through the LL degeneracy.
Due to the spin degeneracy, each shell is further split by the Zeeman 
gap.

Our model applies to the narrow and symmetric QW's, and the calculations 
have been carried out for the GaAs/AlGaAs structures with the Al 
concentration of $x=0.33$ and the widths of $w=10$~nm, 11.5~nm, and 13~nm.
For such systems, only the lowest QW subband need be included and the 
cyclotron motion of both electrons and holes can be well described in 
the effective-mass approximation.\cite{whittaker}
For the holes, only the heavy-hole states are included, with the 
inter-subband coupling partially taken into account through the 
realistic dependence $\hbar\omega_{ch}(B)$, i.e.\ through the 
dependence of the effective in-plane (cyclotron) mass $m_h^*$ on $B$
(after Cole et al.\cite{cole}).

Using a composite index $i=[nm\sigma]$ ($\sigma$ is the spin 
projection), the $e$--$h$ Hamiltonian can be written as
\begin{equation}
\label{eq1}
  H = \sum_{i,\alpha} 
      c_{i\alpha}^\dagger c_{i\alpha} \varepsilon_{i\alpha}
    + \!\!\!\! \sum_{ijkl,\alpha\beta} \!\!\!\!
      c_{i\alpha}^\dagger c_{j\beta}^\dagger c_{k\beta} c_{l\alpha} 
      V^{\alpha\beta}_{ijkl},
\end{equation}
where $c_{i\alpha}^\dagger$ and $c_{i\alpha}$ create and annihilate 
particle $\alpha$ ($e$ or $h$) in state $i$, and $V^{\alpha\beta}_{ijkl}$ 
are the Coulomb matrix elements.

At high magnetic fields, $w$ significantly exceeds $\lambda$ and it is
essential to properly include the effects due to the finite QW width.
Merely scaling all matrix elements $V^{\alpha\beta}_{ijkl}$ by a constant 
factor $\xi(w/\lambda)$ is not enough.
Ideally, the $V^{\alpha\beta}_{ijkl}$ should be calculated for the actual
3D electron and hole wavefunctions.\cite{whittaker}
The ``rod'' geometry used by Chapman et al.\cite{chapman} might be a 
reasonable approximation (for the lowest QW subband), although using 
the same effective rod length for electrons and holes and its arbitrary 
scaling with $B$ leads to an incorrect $B$-dependence of obtained 
results.
In this work we insist on using numerically correct values of 
$V^{\alpha\beta}_{ijkl}$ and calculate them in the following way.
The actual density profile across the QW can be approximated by 
$\varrho(z)\propto\cos^2(\pi z/w^*)$, i.e.\ by replacing the actual 
QW by a wider one, with an infinite potential step at the interface.
This defines the effective widths of electron and hole layers, 
$w_e^*$ and $w_h^*$.
For $w\sim10$~nm, we obtain $w^*\equiv(w_e^*+w_h^*)/2=w+2.5$~nm.
We have checked that the effective 2D interaction in a quasi-2D layer, 
\begin{equation}
   V(r)=\int\! dz\!\int\! dz'
   {\varrho(z)\varrho(z')\over\sqrt{r^2+(z-z')^2}},
\end{equation}
can be well approximated\cite{he} by $V_d(r)=1/\sqrt{r^2+d^2}$
if an effective separation across the QW is taken as $d=w^*/5$.
For a given $d/\lambda$, matrix elements of $V_d(r)$ have been calculated 
analytically and used as $V^{\alpha\beta}_{ijkl}$ in Eq.~(\ref{eq1}).
A small difference between $w_e^*$ and $w_h^*$ is included by additional 
rescaling, $V_{\alpha\beta}(r)=\xi_{\alpha\beta}V(r)$, with $\xi_{\alpha
\beta}^2=\left<\right.z_{eh}^2\left.\right>/\left<\right.z_{\alpha\beta}^2
\left.\right>$.
For $w\sim10$~nm, we obtain $\xi_{ee}=0.94$ and $\xi_{hh}=1.08$, and
for wider QW's, the difference between $w_e^*$ and $w_h^*$ is even smaller.
Note that our treatment of interactions in a quasi-2D layer is different 
from the ``biplanar'' geometry (electrons and holes confined in two 
parallel infinitely thin layers) tested by Chapman et al.\cite{chapman}

The Hamiltonian $H$ is diagonalized numerically in the basis including 
up to five LL's ($n\le4$) for both electrons and holes (note that since 
$l=S+n$, the inter-LL excitations of only one particle have non-zero 
angular momentum and, e.g., do not contribute to the $X$ ground state).
Energies obtained for different values of $2S\le20$ are extrapolated to 
$2S\rightarrow\infty$, i.e.\ to an infinite QW.
The eigenstates are labeled by total angular momentum $L$ and its 
projection $M$, which are related to the good quantum numbers on the 
plane: ${\cal L}_{\rm tot}$, ${\cal L}_{\rm cm}$, and ${\cal L}_{\rm rel}
\equiv{\cal L}_{\rm tot}-{\cal L}_{\rm cm}$.
The total electron and hole spins ($J_e$ and $J_h$) and projections 
($J_{ze}$ and $J_{zh}$) are also resolved.

\section{Bound $X^-$ States}
\begin{figure}[t]
\epsfxsize=3.40in
\epsffile{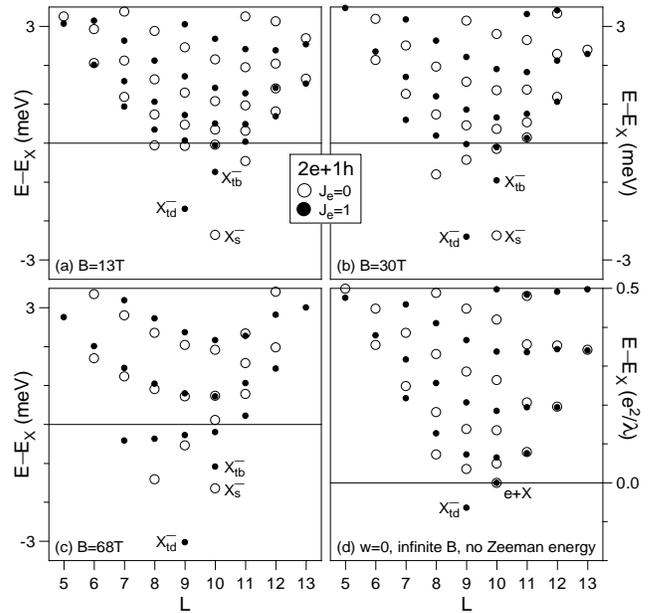}
\caption{
   The energy spectra (energy $E$ vs.\ angular momentum $L$) of the 
   $2e$--$1h$ system on a Haldane sphere with the Landau level degeneracy 
   of $2S+1=21$.
   $E_X$ is the exciton energy.
   The parameters are appropriate for the 11.5~nm GaAs quantum well.
}
\label{fig1}
\end{figure}

The $2e$--$1h$ energy spectra calculated for $2S=20$ and five included 
electron and hole LL's ($n\le4$) are shown in Fig.~\ref{fig1}.
The parameters used in the calculation ($w_e^*$, $w_h^*$, and the 
dependence of $\hbar\omega_{ch}$ on $B$) correspond to the 11.5~nm
GaAs QW.
The energy is measured from the exciton energy $E_X$, so that for the 
bound states (the states below the lines) it is opposite to the binding 
energy $\Delta$ (the lowest LL energy is set to zero).
Open and full symbols denote singlet and triplet electron spin 
configurations, respectively, and only the state with the lowest Zeeman 
energy is marked for each triplet.
Similarly, each state with $L>0$ represents a degenerate multiplet with 
$|M|\le L$.
The angular momentum $L$ calculated in the spherical geometry translates 
into angular momenta on a plane in such way\cite{geometry} that the 
$L=S$ multiplet corresponds to ${\cal L}_{\rm rel}=0$ and ${\cal L}_{\rm 
tot}={\cal L}_{\rm cm}=0$, 1, \dots, and the $L=S-1$ multiplet corresponds 
to ${\cal L}_{\rm rel}=1$ and ${\cal L}_{\rm tot}={\cal L}_{\rm cm}+1=1$, 
2, \dots.

Due to the conservation of $L$ in the PL process, only states from 
the $L=S$ channel are radiative.
This is because\cite{chen,x-fqhe,x-cf} an annihilated $e$--$h$ pair has 
$L_X=0$, and the final-state electron left in the lowest LL has $l_e=S$.
Recombination of other, non-radiative ($L\ne S$) states requires breaking 
rotational symmetry, e.g., by interaction with electrons, other charged 
complexes, or impurities.
This result is independent of chosen spherical geometry and holds also 
for the planar QW's, where the 2D translational symmetry leads to the 
conservation of ${\cal L}_{\rm tot}$ and ${\cal L}_{\rm cm}$, and the 
corresponding PL selection rule is ${\cal L}_{\rm rel}=0$ (this simple 
result can also be expressed in terms of magnetic translations
\cite{dzyubenko2}).

Three states marked in Fig.~\ref{fig1}(a,b,c) ($B=13$, 30, and 68~T)
are of particular importance.
$X^-_s$ and $X^-_{tb}$, the lowest singlet and triplet states at $L=S$, 
are the only well bound radiative states, while $X^-_{td}$ has by far 
the lowest energy of all non-radiative ($L\ne S$) states.
The transition from the $X^-_s$ to the $X^-_{td}$ ground state is found 
at $B\approx30$~T, which confirms the calculation of WS.
Our slightly larger binding energies for $w=10$~nm are due to a larger 
basis used for diagonalization and including the magnetic-field dependence 
of the effective hole cyclotron mass (for $w\sim10$~nm, $m_h^*$ increases 
\cite{cole} from 0.28 at 10~T to 0.40 at 50~T).
A new result is the identification of the $X^-_{tb}$ state, which remains 
an excited radiative bound state in all frames (a)--(c).

For comparison, the spectrum of an ideal, strictly 2D system in the 
lowest LL is shown in Fig.~\ref{fig1}(d).
The $X^-_{td}$ is the only bound state.\cite{x-dot}
As a result of the hidden symmetry,\cite{macdonald,lerner,dzyubenko1,chen} 
the only radiative states are the pair of ``multiplicative'' states at 
$L=S$ and $E=E_X$, in which an optically active $X$ with $L_X=0$ is 
decoupled from a free electron with $l_e=S$.

We have performed similar calculations for systems larger than $2e$--$1h$.
The results confirm that $X$ and $X^-$ are the only well bound $e$--$h$ 
complexes at $B\ge10$~T.
For example, the charge-neutral singlet biexciton $X_2$ (with $J_e=J_h=
L_{X_2}=0$) unbinds at $B\approx20$~T even in the absence of the Zeeman 
splitting, and its Coulomb binding energy between 10 and 20~T is less 
than 0.1~meV.

\begin{figure}[t]
\epsfxsize=3.40in
\epsffile{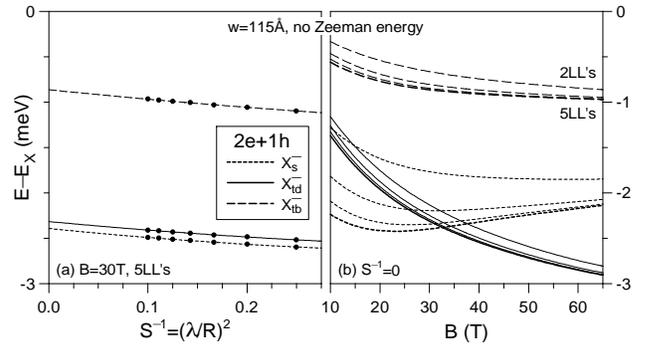}
\caption{
   (a) 
   The $X^-$ energies $E$ calculated on a Haldane with the Landau level 
   (LL) degeneracy $2S+1$, plotted as a function of $S^{-1}$, i.e. the 
   sphere radius $R$.
   $E_X$ is the exciton energy and $\lambda$ is the magnetic length.
   Five electron and hole LL's are included.
   (b) 
   The $X^-$ binding energies $E$ extrapolated to $\lambda/R=0$,
   plotted as a function of the magnetic field $B$.
   Data obtained including different numbers of electron and hole 
   LL's are shown; thicker lines are for five LL's.
   The parameters in both frames are are appropriate for the 11.5~nm
   GaAs quantum well.
}
\label{fig2}
\end{figure}

To illustrate the finite size and surface curvature effects on the 
results obtained in the spherical geometry, in Fig.~\ref{fig2}(a) we 
plot the Coulomb binding energies (without the Zeeman energy) of all 
three $X^-$ states marked in Fig.~\ref{fig1}(b) ($B=30$~T) as a function 
of $S^{-1}=(\lambda/R)^2$.
The very regular dependence of the binding energies on the system size 
allows accurate extrapolation of the values obtained for $8\le2S\le20$ 
to $2S\rightarrow\infty$, i.e.\ to an extended planar system 
($\lambda/R=0$ and infinite LL degeneracy).

The effect of LL mixing is demonstrated in Fig.~\ref{fig2}(b), where we 
plot the extrapolated binding energies ($\lambda/R=0$), calculated 
including between two and five electron and hole LL's, as a function 
of $B$.
The following observations can be made.
Although inclusion of one excited ($n=1$) LL already leads to a 
significant $X^-_s$ binding, at least the $n=2$ level must be added for 
the quantitatively meaningful results.
Because the singlet state has more weight in the excited LL's than the 
triplet states, the ground-state transition shifts to higher $B$ when 
more LL's are included.
The $X^-_s$ binding energy $\Delta_s$ weakly depends on $B$ and saturates 
at $B\approx20$~T, while $\Delta_{td}\propto e^2/\lambda\propto\sqrt{B}$.
Finally, the $X^-_{tb}$ energy goes at a roughly constant separation of 
1.5~meV above $X^-_s$, and never crosses either $X^-_s$ or $X^-_{td}$.

\begin{figure}[t]
\epsfxsize=3.40in
\epsffile{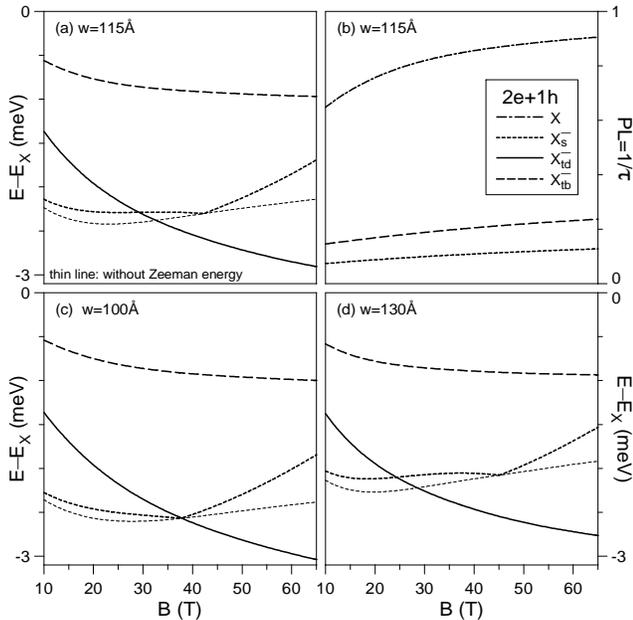}
\caption{
   The $X^-$ binding energies $E$ (acd) and the photoluminescence 
   intensities $\tau^{-1}$ (b) calculated for 10~nm (c), 11.5~nm 
   (ab), and 13~nm GaAs quantum wells, plotted as a function of 
   the magnetic field $B$.
   $E_X$ is the exciton energy.
}
\label{fig3}
\end{figure}

To illustrate the dependence on the QW width, in Fig.~\ref{fig3}(a,c,d) 
we compare the $X^-$ binding energies obtained for $w=10$~nm, 11.5~nm, 
and 13~nm.
The thick dotted lines for $X^-_s$ include the Zeeman energy needed to 
flip one electron's spin and form a bound spin-unpolarized state in an 
initially spin-polarized electron gas.
The Zeeman energy $E_Z=g^*\mu_B B$ is roughly a linear function of 
energy through both cyclotron energy $\hbar\omega_c\propto B$ and 
confinement energy $\propto1/w^2$.
After Snelling et al.,\cite{snelling} for $w\sim10$~nm at $B=0$, we 
have $(g_e^*+0.29)w^2=9.4$~nm$^2$, and after Seck, Potemski, and Wyder
\cite{seck} we find $dg_e^*/dB=0.0052$~T$^{-1}$ (for very high fields 
see also Ref.~\onlinecite{najda}).
In all frames, $E_Z$ changes sign at $B\sim40$~T, resulting in cusps in 
the $X^-_s$ binding energy.

To explicitly show the magnitude of $E_Z$, with thin lines we also plot 
the $X^-_s$ energy without $E_Z$.
While the $\Delta_s$ {\em including} $E_Z$ governs the $X^-_s$ 
relaxation and dependence of the $X^-_s$ PL intensity on temperature, 
the $\Delta_s$ {\em without} $E_Z$ is the difference between the $X$ 
and $X^-_s$ PL energies (neglecting the difference\cite{glasberg} 
between $g^*_h$ in the two complexes).
It is clear from Fig.~\ref{fig3}(a,c,d) that $E_Z$ is almost negligible 
for $B<50$~T and that the binding energies are similar for all three 
widths.

Since only three $e$--$h$ complexes: $X$, $X^-_s$, and $X^-_{tb}$, 
have significant binding energy and at the same time belong to the 
radiative $L=S$ channel, only three peaks are observed in the PL 
spectra of dilute systems (not counting the Zeeman splittings).
The total oscillator strength $\tau_\psi^{-1}$ of a given state 
$\psi$ can be expressed as 
\begin{equation}
\label{eq2}
   \tau_\psi^{-1}=
   \left<\psi|{\cal P}^\dagger{\cal P}|\psi\right>,
\end{equation}
where ${\cal P}^\dagger=\sum_i (-1)^mc^\dagger_{ie}c^\dagger_{ih}$ 
and ${\cal P}=\sum_i (-1)^mc_{ie}c_{ih}$ are the optical operators 
coherently creating and annihilating an $e$--$h$ pair with $L=0$
(optically active $X$).
In Fig.~\ref{fig3}(b), we plot $\tau^{-1}$ of $X$, $X^-_s$, and 
$X^-_{tb}$ as a function of $B$ for the 11.5~nm QW.
The units of $\tau^{-1}$ follow from Eq.~(\ref{eq2}).
We assume here that both electrons and holes are completely 
spin-polarized ($J_z=J$).
Typically, all electron spins and only a fraction of hole spins 
$\chi_h$ (depending on temperature and the Zeeman energy) are aligned 
with the field.
In result, the $X^-_{tb}$ PL has definite circular polarization 
($\sigma_+$) and its intensity is reduced by $\chi_h$, while the 
$X^-_s$ PL peak splits into a $\sigma_\pm$ doublet (separated by 
the appropriate Zeeman energy) with the intensity of the two 
transitions weighted by $\chi_h$ and $1-\chi_h$.

In a system obeying the hidden symmetry ($w_e^*=w_h^*$, no LL mixing,
and no QW subband mixing), the total oscillator strength of one $X$ 
is equally shared by a pair of multiplicative $e$--$X$ states.
In Fig.~\ref{fig3}(b), it is distributed over a number of radiative 
($L=S$) states, and, although most is inherited by the two nearly 
multiplicative states at $E\approx E_X$, a fraction also goes to 
the well bound $X^-_s$ and $X^-_{tb}$ states, with the ratio 
$\tau_{tb}^{-1}\approx2\tau_s^{-1}$ almost independent of $B$.
The resulting three PL peaks ($X$, $X^-_s$, and $X^-_{tb}$) are 
precisely the ones observed in experiments.\cite{shields,finkelstein,%
glasberg,priest,hayne}

The actual relative intensity of the PL peaks will depend not only on 
the oscillator strengths but also on the relative population of the 
respective initial states (i.e., efficiency of the relaxation processes, 
which in turn depends on the excitation energies and temperature) and 
their spin polarization.
An increase of $\chi_h$ from ${1\over2}$ to 1 with increasing $B$ can
explain an increase of the $X^-_{tb}$ PL intensity by up to a factor of 
two, while the $X^-_s$ PL intensity remains roughly constant.\cite{hayne}

Let us stress that the results presented in Figs.~\ref{fig1}--\ref{fig3}
are appropriate for narrow and symmetrically (or remotely) doped QW's.
The agreement of the calculated binding energies and their dependence 
on $B$ with the experimental data for such systems
\cite{shields,finkelstein,glasberg} is good.
In much wider QW's ($w\sim30$~nm), the subband mixing becomes 
significant\cite{whittaker} (and favors the $X^-_s$ ground state), 
while in strongly asymmetric QW's or heterojunctions the Coulomb 
matrix elements $V^{\alpha\beta}_{ijkl}$ are quite different.
In the latter case, the significant difference between electron and 
hole QW confinements ($w_h^*\ll w_e^*$) increases the $e$--$h$ 
attraction compared to the $e$--$e$ repulsion within an $X^-$.
Roughly, the binding energies of all three $X^-$ states increase 
(compared to the values calculated here) by an uncompensated $e$--$h$ 
attraction which scales as $e^2/\lambda\propto\sqrt{B}$.
This most likely explains the origin of an (equal) increase of $\Delta_s$ 
and $\Delta_{tb}$ as a function of $B$ found in Ref.~\onlinecite{hayne}.

While a quantitative model adequate to asymmetric QW's or heterojunctions 
must use correct (sample-dependent) electron and hole charge density 
profiles $\varrho(z)$, our most important result remains valid for all 
structures: 
The triplet $X^-$ state seen in PL is the ``bright'' excited triplet 
state at $L=S$ (${\cal L}_{\rm rel}=0$), while the lowest triplet state 
at $L=S-1$ (${\cal L}_{\rm rel}=1$) so far remains undetected.

It might be useful to realize which of the experimentally controlled 
factors generally shift the singlet-triplet $X^-$ transition to lower 
magnetic fields.
The hidden symmetry which in Fig.~\ref{fig1}(d) prevents binding of any 
other states than $X^-_{td}$ is the exact overlap of electron and 
hole orbitals.\cite{macdonald,lerner,dzyubenko1,chen}
The experimentally observed binding of $X^-_s$ is due to the 
confinement of the hole charge in a smaller volume (through asymmetric 
LL mixing, $\hbar\omega_{ce}>\hbar\omega_{ch}$, and asymmetric QW 
confinement, $w_e^*>w_h^*$), which enhances the $e$--$h$ attraction 
compared to the $e$--$e$ repulsion.
Therefore, any factors should be avoided which break the $e$--$h$ 
orbital symmetry, such as 
(i) large $w$ leading to the QW subband mixing, 
(ii) well/barrier material combinations yielding $w_h^*\ll w_h^*$, 
(iii) large in-plane effective masses and small dielectric constants 
(large $[e^2/\epsilon\lambda]/[\hbar\omega_c]$) leading to the strong 
LL mixing.
On the other hand, reducing $w$ strengthens Coulomb interaction and 
thus LL mixing, while too weak interactions (scaled e.g.\ by $\epsilon$) 
might decrease $X^-$ binding energies below the experimental resolution.
The giant electron Zeeman splitting ($g_e^*\sim1$) in CdTe or ZnSe 
structures\cite{wojtowicz,astakhov} might certainly help to stabilize 
the $X^-_{td}$ ground state at low $B$.
Also, appropriate asymmetric doping producing an electric field across 
the QW and slightly separating electron and hole layers can help to 
restore balance between the $e$--$e$ and $e$--$h$ interactions.
 
\section{Effects of $X^-$ Interactions}
Even in dilute systems, recombination of bound $e$--$h$ complexes can 
in principle be affected by their interaction with one another or with 
excess electrons.
The short-range part of the $e$--$X^-$ and $X^-$--$X^-$ interaction 
potentials is weakened due to the $X^-$ charge polarization,
\cite{x-fqhe,x-cf} and it is not at all obvious if even in the narrow 
QW's the resulting $e$--$X^-$ and $X^-$--$X^-$ correlations will be 
similar to the Laughlin\cite{laughlin} correlations in an electron gas.
Instead, the long-range part of the effective potentials could lead to 
some kind of $e$--$X^-$ or $X^-$--$X^-$ pairing (in analogy to the 
electron or composite Fermion pairing\cite{haldane2} in the fractionally 
filled excited LL's).
It has been shown\cite{parentage} that the repulsion has short range 
and results in Laughlin correlations, if its pseudopotential $V(L)$,
defined\cite{haldane2} as the pair interaction energy $V$ as a function 
of the pair angular momentum $L$ (on a sphere, larger $L$ means smaller 
separation) increases more quickly than $L(L+1)$.
Therefore, the correlations in an infinite system (QW) are determined
by form of the relevant pseudopotentials which can be obtained from 
studies of relatively small systems.

\begin{figure}[t]
\epsfxsize=3.40in
\epsffile{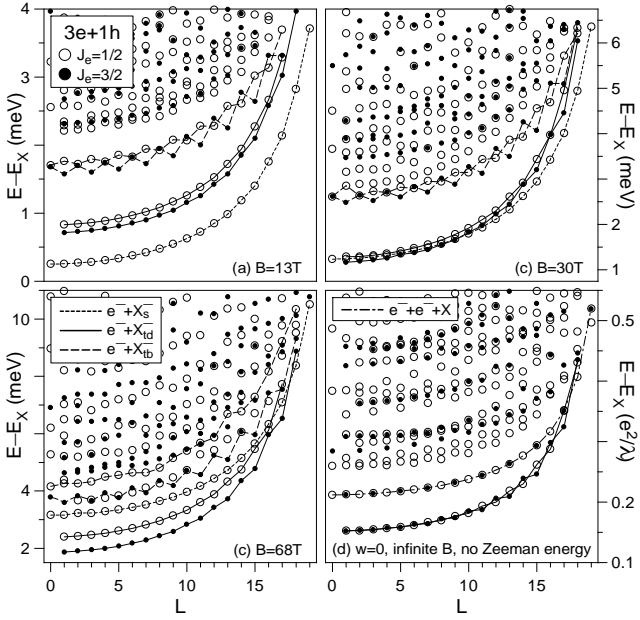}
\caption{
   The energy spectra (energy $E$ vs.\ angular momentum $L$) of the 
   $3e$--$1h$ system on a Haldane sphere with the Landau level degeneracy 
   of $2S+1=21$.
   $E_X$ is the exciton energy.
   The parameters in frames (abc) are appropriate for the 11.5~nm
   GaAs quantum well.
   In frame (d), $\lambda$ is the magnetic length.
}
\label{fig4}
\end{figure}

In Fig.~\ref{fig4} we plot the energy spectra of an $3e$--$1h$ system
(the simplest system in which an $X^-$ interacts with another charge), 
calculated for $2S=20$ and three electron and hole LL's included 
($n\le2$).
The open and filled circles mark the states with total electron spins 
$J_e={1\over2}$ and ${3\over2}$, respectively, and only the lowest 
energy states are shown for each spin multiplet.
In the low energy states, bound $X^-$ complexes interact with an 
electron through the effective pseudopotentials $V(L)$, and the total 
energy of an interacting pair is the sum of $V(L)$ and $E_{X^-}$.
For each pair, the allowed $L$ are obtained by adding $l_e=S$ of an 
electron and $L_{X^-}$ of an appropriate $X^-$.
This yields $L\ge0$ for $X^-_s$ and $X^-_{tb}$, and $L\ge1$ for $X^-_{td}$.
However, maximum $L$ are smaller than $L_{X^-}+S$ due to the finite size
of the $X^-$ (hard core).\cite{x-fqhe,x-cf}
The allowed total electron spins $J_e$ are obtained by adding ${1\over2}$ 
of an electron to 0 or 1 of an $X^-$, so that the $e$--$X^-_s$ pair 
states have $J_e={1\over2}$, while the $e$--$X^-_{tb}$ and $e$--$X^-_{td}$ 
pair states can have either $J_e={1\over2}$ or ${3\over2}$.

At low $L$ (i.e., at low $e$--$X^-$ interaction energy compared to the 
$X^-$ binding energy), the $e$--$X^-$ scattering is decoupled from 
internal $X^-$ dynamics, and all $e$--$X^-$ pseudopotentials marked 
with lines in Fig.~\ref{fig4} are rather well approximated by those 
of two distinguishable point charges (electrons) with appropriate $l$'s.
Their relative position in different $e$--$X^-$ bands depends on involved 
$\Delta$ and $E_Z$, and the $e$--$X^-_{td}$ states form the lowest energy 
band at sufficiently large $B$; see Fig.~\ref{fig4}(c).
Such regular behavior of the (two-charge) $3e$--$1h$ system implies 
\cite{x-fqhe,x-cf} that the lowest states of an infinite $e$--$h$ plasma
are formed by bound $X^-$'s interacting with one another and with excess 
electrons through the Coulomb-like pseudopotentials.
Depending on $B$, either $X^-_s$'s or $X^-_{td}$'s form the ground state, 
while other bound complexes occur at higher energies, with the excitation 
gap given by the appropriate difference in $\Delta$ and $E_Z$.

Less obviously, because of the short-range\cite{parentage} character of 
$V(L)$, the low-lying states have Laughlin--Halperin\cite{laughlin,halperin} 
$e$--$X^-$ correlations described by a Jastrow prefactor 
$\prod(x_i-y_j)^\mu$, where $x$ and $y$ are complex coordinates of 
electrons and $X^-$'s, respectively, and $\mu$ is an integer.
At fractional LL fillings $\nu=\nu_e-\nu_h$, $X^-$'s avoid\cite{parentage} 
as much as possible the $e$--$X^-$ pair states with largest values of $L$.
At $\nu=1/\mu$, the ground state is the Laughlin-like incompressible fluid 
state with $L\le L_{X^-}+S-\mu$, with quasiparticle-like excitations 
described by a generalized composite Fermion model.\cite{x-fqhe,x-cf}
Even though formation of an equilibrium $X^-$ Laughlin state requires 
long $X^-$ lifetime and hence is only likely for $X^-_{tb}$, all $X^-$'s 
will stay as far as possible from other charges, and the distance to the 
nearest one corresponds to the $L=L_{X^-}+S-\mu$ pair state.
This result depends on our assumption of the small QW width $w$.
He at al.\cite{he} showed that the Laughlin $e$--$e$ correlations are 
destroyed in a thick GaAs QW when $w/\lambda>6$.
At $B=40$~T, this corresponds to $w>24$~nm, but the critical width for 
the $e$--$X^-$ correlations will be even smaller because of the above 
mentioned $X^-$ charge polarization\cite{x-fqhe,x-cf} 

\begin{figure}[t]
\epsfxsize=3.40in
\epsffile{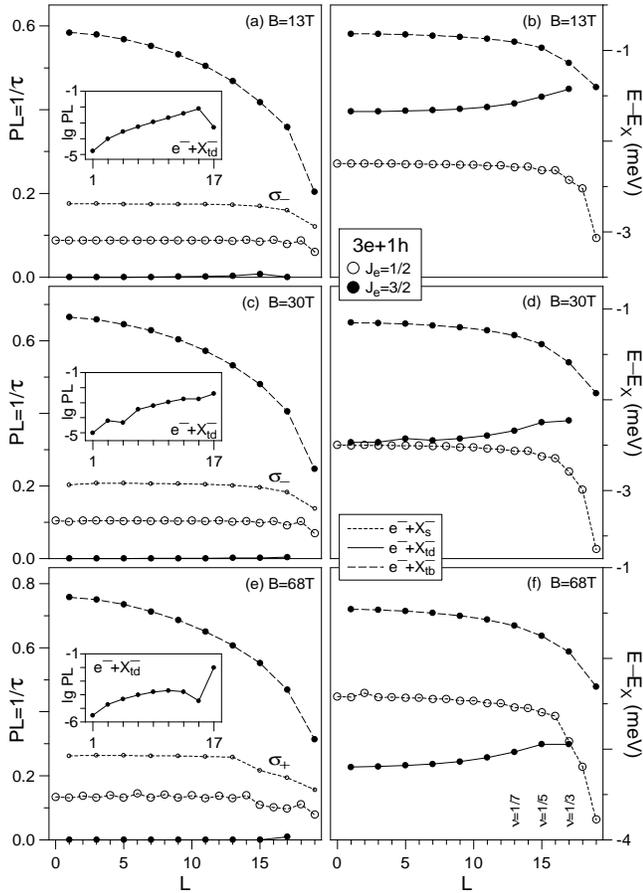}
\caption{
   The photoluminescence intensities $\tau^{-1}$ (left) and energies $E$ 
   (right) of an $X^-$ interacting with an electron on a Haldane sphere 
   with the Landau level degeneracy of $2S+1=21$, plotted as a function 
   of the $e$--$X^-$ pair angular momentum $L$.
   $E_X$ is the exciton energy.
   The parameters are appropriate for the 11.5~nm GaAs quantum well.
}
\label{fig5}
\end{figure}

The connection between $\nu$ and the minimum allowed $e$--$X^-$ separation 
(or $L$) allows calculation of the effect of the $e$--$X^-$ interaction on 
the $X^-$ recombination as a function of $\nu$.
In Fig.~\ref{fig5} we plot the PL oscillator strength $\tau^{-1}$ and 
energy $E$ (measured from the exciton energy $E_X$) for some of the 
$3e$--$1h$ states marked in Fig.~\ref{fig4}(a,b,c).
We assume that the Zeeman energy will polarize all electron spins prior 
to recombination, except for those two in the $X^-_s$, and concentrate 
on the following three initial configurations: 
$e$--$X^-_s$ with $J_{ze}=J_e={1\over2}$ and
$e$--$X^-_{tb}$ and $e$--$X^-_{td}$ with $J_{ze}=J_e={3\over2}$.
For each of the three configurations, $\tau^{-1}$ and $E$ are plotted 
as a function of $L$ (i.e.\ of $\nu$).

The quantities conserved in an emission process are the total angular 
momentum $L$ and its projection $M$ (on a plane, ${\cal L}_{\rm tot}$ 
and ${\cal L}_{\rm cm}$), and the total electron and hole spins and 
their projections change by $\pm{1\over2}$.
For $X^-_{tb}$ and $X^-_{td}$, only an $e\!\!\uparrow$--$h\!\!\downarrow$ 
pair can be annihilated, and an emitted photon has a definite circular 
polarization $\sigma_+$.
Two indistinguishable electrons left in the final state have the total
spin $J_e=1$, so their pair angular momentum $L$ must be odd ($2l_e$ 
minus an odd integer).
For $X^-_s$, both $\sigma_+$ and $\sigma_-$ PL are possible, with the
energy of the latter transition shifted by the total electron and hole 
Zeeman energy.
For $\sigma_+$, the two electrons in the final state can have either 
$J_e=0$ and $L$ even, or $J_e=1$ and $L$ odd; while for $\sigma_+$ they 
can only have $J_e=1$ and $L$ must be odd.
Note that $g_e^*$ changes sign at $B\approx42$~T, and the polarizations
in Fig.~\ref{fig5}(e) are reversed.
As expected, for $L\rightarrow0$ the oscillator strengths converge to 
those of appropriate single $X^-$'s in Fig.~\ref{fig3}(b) (multiplied 
by two if only one parity of $L$ is allowed).
On the right-hand side of Fig.~\ref{fig5}, the $\sigma_+$ PL energies 
are shown.
For only partial polarization of hole spins, an unmarked $\sigma_-$ peak 
of an $X^-_s$ will appear at the energy higher by the $X^-$ (not electron) 
Zeeman splitting.\cite{glasberg}

There is no significant effect of the $e$--$X^-$ interactions on the 
$X^-$ oscillator strength and energy at small $L$.
Moreover, the decrease of the PL energy of an $X^-_s$ at larger $L$ is 
due to its induced charge polarization (dipole moment).\cite{x-fqhe,x-cf} 
This effect is greatly reduced for an $X^-$ surrounded by an isotropic 
electron gas, although slight residual variation of the PL energy at 
$\nu\sim{1\over3}$ might broaden the $X^-_s$ peak.
The insensitivity of the $X^-$ recombination to the $e$--$X^-$ 
interactions at small $L$ justifies a simple picture of PL in dilute 
$e$--$h$ plasmas.
In this picture, recombination occurs from a single isolated bound 
complex and hence is virtually insensitive\cite{priest} to $\nu$.
Quite surprisingly, the LH correlations prevent increase of the 
$X^-_{td}$ oscillator strength through interaction with other charges.
$\tau^{-1}_{td}$ decreases exponentially (see insets) with decreasing 
$\nu$, and remains ten times longer than $\tau_s$ even at $\nu={1\over3}$.
This explains the absence of an $X^-_{td}$ peak even in the PL spectra 
\cite{shields,finkelstein,glasberg,priest,hayne} showing strong 
recombination of a higher-energy triplet state $X^-_{tb}$ (however, 
see also Ref.~\onlinecite{munteanu}).

\section{Conclusion}
We have studied photoluminescence (PL) from a dilute 2D electron gas 
in narrow and symmetric quantum wells (QW's) as a function of the 
magnetic field $B$ and the QW width.
The puzzling qualitative discrepancy between experiments and earlier 
theories is resolved by identifying the radiative ($X^-_{tb}$) and 
non-radiative ($X^-_{td}$) bound states of a triplet charged exciton.
Even in high magnetic fields, when it has lower energy than the 
radiative states, the $X^-_{td}$ remains invisible in PL experiments
due to its negligible oscillator strength.
The short range of the $e$--$X^-$ interaction pseudopotentials results
in the Laughlin--Halperin correlations in a dilute $e$--$h$ plasma,
and effectively isolates the bound $X^-$ states from the remaining 
electrons.
This explains the observed insensitivity of the PL spectra to the 
filling factor and persistence of the small $X^-_{td}$ oscillator 
strength in an interacting system.
An idea of the Laughlin incompressible-fluid states of long-lived 
$X^-_{td}$'s is supported.
The ``dark'' $X^-_{td}$ state could be identified either in 
time-resolved PL or transport experiments.

\section*{Acknowledgment}
The authors wish to thank M. Potemski (HMFL Grenoble) and C. H. Perry
and F. Munteanu (LANL Los Alamos) for helpful discussions.
AW and JJQ acknowledge partial support by the Materials Research 
Program of Basic Energy Sciences, US Department of Energy.

\end{document}